\documentclass[sigconf]{acmart} 

\AtBeginDocument{%
  \providecommand\BibTeX{{%
    \normalfont B\kern-0.5em{\scshape i\kern-0.25em b}\kern-0.8em\TeX}}}

\copyrightyear{2021} 
\acmYear{2021} 
\setcopyright{acmcopyright}
\acmConference[MM '21]{Proceedings of the 29th ACM International Conference on Multimedia}{October 20--24, 2021}{Virtual Event, China}
\acmBooktitle{Proceedings of the 29th ACM International Conference on Multimedia (MM '21), October 20--24, 2021, Virtual Event, China}
\acmPrice{15.00}
\acmDOI{10.1145/3474085.3479238}
\acmISBN{978-1-4503-8651-7/21/10}

\usepackage[linesnumbered,ruled,lined]{algorithm2e}
\usepackage{balance}
\usepackage{multirow}
\usepackage{dblfloatfix}
\usepackage{bbding}
\usepackage{subcaption}
\usepackage{makecell}
\usepackage{array}
\usepackage{booktabs}

\begin{document}
\fancyhead{}
\title{Metaverse for Social Good: A University Campus Prototype}

\author{Haihan Duan}
\affiliation{%
  \institution{The Chinese University of Hong Kong, Shenzhen\\
  Shenzhen Institute of Artificial Intelligence and Robotics for Society}
  \city{Shenzhen}
  \country{China}}
\email{haihanduan@link.cuhk.edu.cn}

\author{Jiaye Li}
\affiliation{%
  \institution{The Chinese University of Hong Kong, Shenzhen\\
  Shenzhen Institute of Artificial Intelligence and Robotics for Society}
  \city{Shenzhen}
  \country{China}}
\email{jiayeli@link.cuhk.edu.cn}

\author{Sizheng Fan}
\affiliation{%
  \institution{The Chinese University of Hong Kong, Shenzhen\\
  Shenzhen Institute of Artificial Intelligence and Robotics for Society}
  \city{Shenzhen}
  \country{China}}
\email{sizhengfan@link.cuhk.edu.cn}

\author{Zhonghao Lin}
\affiliation{%
  \institution{The Chinese University of Hong Kong, Shenzhen\\
  Shenzhen Institute of Artificial Intelligence and Robotics for Society}
  \city{Shenzhen}
  \country{China}}
\email{zhonghaolin@link.cuhk.edu.cn}

\author{Xiao Wu}
\affiliation{%
  \institution{White Matrix Inc.}
  \city{Nanjing}
  \country{China}}
\email{wuxiao@whitematrix.io}

\author{Wei Cai}
\authornote{Wei Cai is the corresponding author (caiwei@cuhk.edu.cn).}
\affiliation{%
  \institution{The Chinese University of Hong Kong, Shenzhen\\
  Shenzhen Institute of Artificial Intelligence and Robotics for Society}
  \city{Shenzhen}
  \country{China}}
\email{caiwei@cuhk.edu.cn}

\renewcommand{\shortauthors}{Duan, et al.}
\fancyhead{} 

\begin{abstract}
In recent years, the metaverse has attracted enormous attention from around the world with the development of related technologies. The expected metaverse should be a realistic society with more direct and physical interactions, while the concepts of race, gender, and even physical disability would be weakened, which would be highly beneficial for society. However, the development of metaverse is still in its infancy, with great potential for improvement. Regarding metaverse’s huge potential, industry has already come forward with advance preparation, accompanied by feverish investment, but there are few discussions about metaverse in academia to scientifically guide its development. In this paper, we highlight the representative applications for social good. Then we propose a three-layer metaverse architecture from a macro perspective, containing infrastructure, interaction, and ecosystem. Moreover, we journey toward both a historical and novel metaverse with a detailed timeline and table of specific attributes. Lastly, we illustrate our implemented blockchain-driven metaverse prototype of a university campus and discuss the prototype design and insights.
\end{abstract}




\maketitle
\section{Introduction}\label{sec_introduction}

The term ``metaverse" originates from the science fiction novel, \emph{Snow Crash} \cite{stephenson2014snow}, written by Neal Stephenson. Metaverse is a combination of ``meta" (meaning beyond) and the stem ``verse" from "universe", denoting the next-generation Internet in which the users, as avatars, can interact with each other and software applications in a three-dimensional (3D) virtual space. There has been approximately 30 years’ development behind the evolution of this term. In 2018, the success of the film \emph{Ready Player One} \cite{spielberg2018ready} brought the concept of metaverse back to the forefront of cutting-edge discussions. This film describes a virtual world named "OASIS", in which everyone could connect to the virtual world, act as their own customized avatar, and do everything they wanted to, based on the basic rules. The film \emph{Ready Player One} shows many seemingly reachable technologies, e.g. head-mounted display (HMD) for virtual reality (VR) rendering, ubiquitous sensing, haptic feedback, and modeling of the physical world, which brings the public feasible opportunities to achieve the metaverse.

However, progress in the development of multimedia technologies (VR/AR, etc.) alone cannot solve all of the problems in the development of the metaverse, since multimedia technologies cannot ensure the digital economy is transparent, stable, and sustainable. For example, the current digital economy is maintained by centralized operators (e.g., large companies), which means that digital properties actually belong to the operators, rather than the users. Fortunately, recent explosive blockchain-related technologies \cite{berg2019blockchain} can be helpful. First proposed by Chaum \cite{chaum1979computer}, blockchain applications in tamper-resistant decentralized ledgers \cite{nofer2017blockchain} have attracted tremendous attention due to the success of Bitcoin \cite{nakamoto2019bitcoin}. In 2013, Vitalik Buterin proposed a decentralized platform named Ethereum \cite{buterin2014next}, which introduces a smart contract for autonomous and transparent program execution, with thousands of novel decentralized applications (DApps) developed \cite{cai2018decentralized}, e.g. blockchain games \cite{min2019blockchain}. Specifically, decentralized finance (DeFi) could ensure that digital properties are unique, persistent, and tradable. Evidently, the metaverse, as an interactive multimedia community relying on massive numbers of online users, may benefit from the technical advances of blockchain to build a fair, free and sustainable society.

Realistic demands and the prospect of feasibly constructing the metaverse motivate the industry, which has been working hard to prepare with fervor. For example, The Sandbox\footnote{https://www.sandbox.game/} is a blockchain-based virtual sandbox game that obtained more than \$2 million in financing in 2020. Similarly, Roblox\footnote{https://www.roblox.com/}, an online game platform and creation system, reached more than \$40 billion in value at its peak. For industrial applications, NVIDIA has built a platform named Omniverse to support real-time virtual collaboration in industrial design and visualization. In addition, there are many giant companies that are preparing to join the metaverse venture. Facebook purchased VR device manufacturer Oculus for deeper development of VR and augmented reality (AR) technologies. Epic Games claimed they raised \$1 billion to put toward building the metaverse, and Sony also invested \$200 million to support Epic’s vision. Moreover, many giants (Tencent, Bytedance) have paid a great deal of attention to constructing metaverse-related applications.

The exploding focus and investments in metaverse from industry would speed up the development and breakthrough of related technologies, but this rapid progress also leads to many problems. Thus, academia has a responsibility to study related problems and give advice to instruct on the development of the metaverse. In recent years, some papers focused on the metaverse have been made available to the public. In 2013, Dionisio et al. \cite{dionisio20133d} published a survey about 3D virtual worlds and the metaverse. This paper describes five phases of virtual world development and specifies four features of a viable metaverse, including realism, ubiquity, interoperability, and scalability, and looks forward to the future evolution of each phase. In 2016, Nevelsteen \cite{nevelsteen2018virtual} sampled a list of technologies using grounded theory to classify technologies that implement a virtual world, and provided a detailed definition of a "virtual world". The above-mentioned two papers clearly explain the concept of the metaverse and its relationship with the virtual world, and guide subsequent research on the metaverse.

However, with rapid technology development, especially block-chain-related technology, many novel metaverse applications have matured, but are not included in the existing surveys. On the other hand, there is a lack of related discussion about the metaverse architecture. Therefore, focusing on the above situation, we summarize the main contributions of this paper as follows:

% However, with the rapid development of technologies, especially blockchain, many novel metaverse applications have grown up, which are not included by the existing surveys. On the other hand, there lacks relative discussion about the architecture of the metaverse. Therefore, focusing on the above situation, we summarize the main contributions of this paper as follows:

\begin{itemize}
    \item We propose a three-layer architecture of the metaverse from a macro point of view, including (from bottom to top): infrastructure, interaction, and ecosystem. Specifically, we introduce the decentralized ecosystem based on blockchain as a novel trend in metaverse development.
	\item We raise open research questions based on the proposed architecture of metaverse that are imperative to solve.
    \item We review existing novel and representative metaverse applications, most of which are not included in previous surveys, and systematically conclude their attributes.
	\item We implement a blockchain-driven metaverse prototype of The Chinese University of Hong Kong, Shenzhen (CUHKSZ), in which we are continuously constructing the system and conducting user studies for further research.
\end{itemize}

% In the rest of this paper, we propose the three-layer architecture of the metaverse with a set of open research questions in Section \ref{sec_architecture}. Then we review the novel and representative metaverse applications and discuss their attributes in Section \ref{sec_related}. We list some applications for social good in Section \ref{sec_social}. In Section \ref{sec_metaverse}, we illustrate the implemented metaverse prototype and introduce its design and functions. Finally, we conclude this paper in Section \ref{sec_conc}.

\section{Metaverse for Social Good} \label{sec_social}

Although the metaverse is a virtual world, regarded as human-centered computing, it indeed shows a significantly positive impact on the real world, especially in terms of accessibility, diversity, equality, and humanity. In this section, we list some representative applications that reflect metaverse for social good.

\subsection{Accessibility}

Currently, global communication and cooperation among countries has become more and more frequent with the rise of globalization, but geographical distance is an objective obstacle that would increase costs during the process. Moreover, influenced by the COVID-19 pandemic, many events are suspended due to pandemic prevention requirements. However, the metaverse could provide great accessibility to serve different social requirements. For example, many events have been converted to virtual form, supported by the metaverse. In 2020, UC Berkeley held its graduation ceremony on \textit{Minecraft}\footnote{https://www.minecraft.net}. Moreover, on \textit{Fortnite}\footnote{https://www.epicgames.com/fortnite/en-US/home}, there are many virtual events held every day, such as a Travis Scott concert. According to the above-mentioned examples, the metaverse has already become an extension of our daily lives, which could satisfy our social requirements, with lower costs and higher security.

\subsection{Diversity}

Restricted to physical limitations (such as geography, language, etc.), the real world cannot integrate various elements in one place to satisfy the requirements of different people. However, the metaverse has unlimited extension space and seamless scene transformation, which could effectively achieve diversity. There are various interesting scenarios that can be held in the metaverse. For example, \textit{Animal Crossing}\footnote{https://animal-crossing.com/} held a presidential campaign for Joe Biden, and students at Stanford University exhibited their posters in \textit{Second Life}\footnote{https://secondlife.com/}. Not limited by the above examples, various activities can be found in the metaverse, such as education, shopping, political campaign, artwork, pets, haunted houses, etc. Therefore, the diversity requirements of physical society have been greatly satisfied.

% Education, Shopping, Political campaign, Artwork, Pets, Haunted, Adults Contents, etc. 
%Second Life
%Animal Crossing

\subsection{Equality}\label{sec_equality}

% \textcolor{red}{Equality may be better than democracy, we can still use words like``democratic'' in the paragraph.}
% Governance tokens: Decentraland, Axie Infinity

Equality is a spiritual pursuit for human beings, but, in reality, there are many factors that influence equality, such as race, gender, disability, and property. In metaverse, everyone can control customized avatars and exercise their power to build a fair and sustainable society. For example, as an autonomous ecosystem, the metaverse includes an attribute of democracy, allowing participants to maintain order and normal operation. In \textit{Decentraland}\footnote{https://decentraland.org/}, there is a Decentralized Autonomous Organization (DAO) where users could propose and vote on the policies created to determine how the world behaves (e.g., what kinds of wearable items are allowed). The \textit{Axie Infinity}\footnote{https://axieinfinity.com/}, a Pokémon-inspired universe where anyone can earn tokens through skilled gameplay and contributions to the ecosystem, also introduces a decentralized organization mechanism, in which Axie Infinity Shards (AXS) holders can stake their tokens through a staking dashboard and participate in governance votes.

% Equality is the spiritual pursuit for human beings, but, in a realistic society, there are many factors that influence equality, like race, gender, disability, and property. In metaverse, everyone can control customized avatars and exercise their power to build a fair and sustainable society. For example, as an autonomous ecosystem, the metaverse has an attribute of democracy for participants to maintain order and normal operation. In Decentraland, there is a Decentralized Autonomous Organization (DAO) where users could propose and vote on the policies created to determine how the world behaves (e.g., what kinds of wearable items are allowed). The Axie Infinity\footnote{https://axieinfinity.com/}, a Pokémon-inspired universe where anyone can earn tokens through skilled gameplay and contributions to the ecosystem, also introduces a decentralized organization mechanism, in which Axie Infinity Shards (AXS) holders can stake their tokens through the staking dashboard and participate in governance votes. 

\subsection{Humanity}

Humanistic spirit is a kind of universal human self-care, which is manifested in the maintenance, pursuit, and concern for human dignity, value, and destiny. In society, humanity cherishes various spiritual and cultural phenomena left behind by previous generations as a legacy for humanity. The metaverse could be an excellent approach for cultural communication and protection. For example, the metaverse has included cultural relics protection. Experiencing years of weathering, cultural relics in the physical world are fragile and likely easily broken by manmade damage or natural disasters. The Notre Dame de Paris caught fire in 2019, sustaining serious damage to the cathedral’s wooden sections. Fortunately, Ubisoft reconstructed the Notre Dame de Paris as a digital 3D model in \textit{Assassin’s Creed Unity}, which will be utilized to support its reconstruction. In China, the National architects \& Cthuwork have made great efforts to reconstruct Chinese cultural relics as 3D voxel models in \textit{Minecraft}, containing buildings, like the Forbidden City, and famous paintings, such as Qingming Shanghe Tu (Ascending the River at Qingming Festival). In the metaverse, the digital reconstruction of cultural relics not only takes place anywhere in the world, but can also provide evidence for relic restoration.

% Humanistic spirit is a kind of universal human self-care, which is manifested in the maintenance, pursuit, and concern for human dignity, value, and destiny. In society, humanity highly cherishes various spiritual and cultural phenomena leftover by human beings. The metaverse could be an excellent approach for cultural communication and protection. For example, the metaverse has worked in cultural relics protection. Experiencing years of weather-beaten, the cultural relics in the physical world are fragile and likely broken by man-made damage or natural disaster. The Notre Dame de Paris caught fire in 2019, which sustained serious damage to the wooden parts. Fortunately, Ubisoft\footnote{https://www.ubisoft.com/en-us/} reconstructed the Notre Dame de Paris as a digital 3D model in \textit{Assassin’s Creed Unity}, which will be utilized to support the reconstruction by the government of France. In China, the National architects \& Cthuwork have made great efforts to reconstruct Chinese cultural relics as 3D voxel models in Minecraft, containing buildings, like Forbidden City, and famous paintings, such as Qingming Shanghe Tu (Ascending the River at Qingming Festival). In the metaverse, the digital reconstruction of cultural relics can not only be traveled from everywhere around the world, but also provide evidence for relics restoration.

\section{Architecture of the Metaverse} \label{sec_architecture}

\begin{figure}[!h]
	\centering
	\includegraphics[width=1.0\columnwidth]{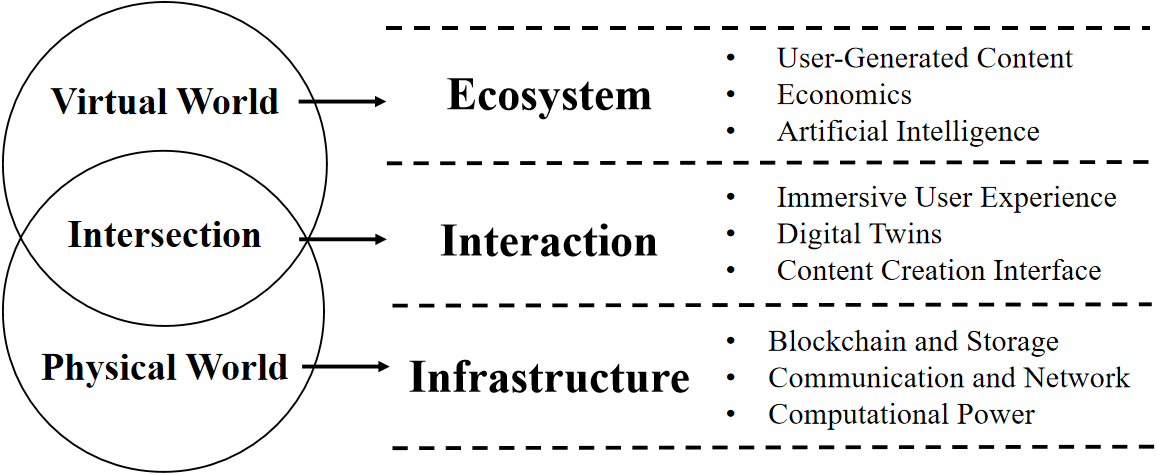}
	\caption{Three-layer Architecture of the Metaverse}
	\label{fig_arc}
\end{figure}

Currently, the development of metaverse is still in the early stages, so its architecture does not have a consistent definition in either academia or industry. For example, Jon Radoff proposed a seven-layer metaverse architecture, where the layers from bottom to top are: infrastructure, human interface, decentralization, spatial computing, creator economy, discovery, and experience \cite{jon2021the}. This architecture is built from an industrial division based on the value chain of the expected market. In contrast, we intend to conclude the architecture of metaverse from a more macro perspective, so a three-layer architecture is proposed, as shown in Figure \ref{fig_arc}, including (from bottom to top): infrastructure, interaction, and ecosystem.

% Currently, the development of metaverse is still in the early stage, so its architecture does not have a consistent definition both in academy and industry. For example, Jon Radoff proposed a seven-layer architecture of the metaverse, where the layers from bottom to top are: infrastructure, human interface, decentralization, spatial computing, creator economy, discovery, and experience \cite{jon2021the}. This architecture is built from an industrial perspective which is a detailed division based on the value chain of the expected market. By contrast, we intend to conclude the architecture of metaverse from a more macro perspective, so a three-layer architecture is proposed in this paper, as shown in Figure \ref{fig_arc}, including (from bottom to top): infrastructure, interaction, and ecosystem.

For the metaverse’s fundamental requirements, we consider that the architecture should cross from the physical world to the virtual world. In Figure \ref{fig_arc}, the left two circles denote the virtual world and physical world, and there is an intersection between the two worlds. The structure of these two circles corresponds to three layers in the central part of Figure \ref{fig_arc}, named infrastructure, interaction, and ecosystem from bottom to top. And we list some key components of each layer right beside the layer name in Figure \ref{fig_arc}. Note that the key components listed will be discussed in this paper, but there are other components in each layer that we will not emphasize, due to this paper’s limited purpose and scope. The proposed three-layer architecture of metaverse can effectively embrace the components that constitute a metaverse, and the details and open research questions of each layer will be discussed further in the following subsections.

\subsection{Infrastructure}

The infrastructure layer contains the fundamental requirements for supporting the operation of a virtual world, including computation, communication, blockchain and storage.

\textbf{Computation and communication.} The metaverse is a large-scale multimedia system, so its operation requires huge computational costs. On the other hand, since the expected metaverse should be accessible at any time and any place, communication technologies are another basic support. In fact, improvements in computation and communication are always cutting-edge research topics, which is not limited to the metaverse. Therefore, in this paper, we pay more attention to how the development of computation and communication could enhance the user experience of the metaverse. We summarize the following open research questions for developing computation and communication for the metaverse as follows: \textit{(1) How to design specific computational devices to support the huge computational consumption of the metaverse? (2) How to coordinate different computational resources, such as cloud computing or mobile devices, to enhance the user experience in different terminals? (3) What kind of data structure or encoding method could effectively present and transmit large-scale data of the metaverse?}

% \textit{(1) How to design specific computational devices to support the high-capacity computation consumption of the metaverse? (2) What kind of data structure or encoding method could effectively present and transmit large-scale multimedia data?}

% \textit{(1) How to design specific computational devices to support the high-capacity computation consumption of the metaverse? (2) How to coordinate different computational resources, such as cloud computing or mobile devices to enhance the user experience in different terminals? (3) What kind of data structure or encoding method could effectively present and transmit large-scale multimedia data? (4) What kind of communication technology could afford the real-time multimedia data transmission of the metaverse?}

\textbf{Blockchain and storage.} The expectations for the metaverse are that it will connect everyone around the world, so an enormous amount of data, such as maps, roles, etc., would be generated and stored in mass storage, which is another basic infrastructure. More importantly, to guarantee decentralization and fairness, the blockchain must be introduced to support sustainable ecosystem operation in the metaverse \cite{cai2018decentralized, berg2019blockchain}. Advanced blockchain systems, like Ethereum \cite{buterin2014next}, introduce a smart contract that could support the operation of DApps, which greatly extend the blockchain’s application scope, and make it feasible for metaverse to construct a decentralized social ecosystem \cite{cai2018decentralized}. In this paper, we regard the blockchain as an infrastructure component of the metaverse, since the essence of blockchain is a distributed ledger, a special kind of storage \cite{chaum1979computer}, and the detailed usability of blockchain in the ecosystem layer will be introduced in Section \ref{sec_ecosystem}. We list the following open research questions that should be considered when building the metaverse: \textit{(1) How to effectively store and retrieve the enormous amount of data in the metaverse? (2) What consensus model should the blockchain adopt to support the sustainable economics of the metaverse? (3) How to reasonably allocate and coordinate data stored in mass storage and the blockchain?}

\subsection{Interaction} \label{sec_interaction}

In this subsection, we emphasize the immersive user experience, digital twins, and content creation, which are important parts in the interaction layer that bridges the physical and virtual worlds. 

\textbf{Immersive user experience.} To achieve an immersive user experience, there are two main components that should be considered in interactions between users and the metaverse. First, the metaverse should receive data from the physical world so that users could control their avatars to finish corresponding actions. As shown in the film \emph{Ready Player One} \cite{spielberg2018ready}, the lead character stands on a treadmill-like machine and wears HMD, gloves, and a special suit so that all user actions could be captured. Second, real-time 3D rendering-related technologies like VR/AR are regarded as the main interaction interface. Moreover, haptic feedback is also necessary, which has already equipped most game controllers, like Nintendo Switch\footnote{https://www.nintendo.com/switch/}. However, existing technologies can only support specific areas, but cannot provide an immersive user experience, which formulates the following research questions: \textit{(1) How to understand the affections of users and enhance their experience during interactions with the metaverse? (2) How to integrate the input and output modalities to build a holistic user experience during interactions?}

\textbf{Digital twins.} With the exception of metaverse users, other objects or things in the physical world could also interact with the metaverse, presented as digital twins \cite{el2018digital} in the virtual world. The parameters of physical devices can be collected by ubiquitous sensing technologies to maintain the same states as their corresponding digital twins. This is an interdisciplinary area that should cover a large number of related subjects, such as material science, signal processing, Internet of Things (IoT), pattern recognition, etc. \cite{essa2000ubiquitous, paulovich2018future}. On the contrary, after operation and processing in metaverse, the parameters in virtual environments can be sent back to physical devices and their real world states can be changed. However, the development of digital twins in the metaverse is still in its early stages, and the following research questions deserve further study: \textit{(1) What in the physical world should be mapped as digital twins within the metaverse? (2) How to use digital twins in the metaverse to effectively benefit the real world?}

% \textbf{Digital twins.} Except for the users of the metaverse, other objects or things in the physical world could also interact with the metaverse, which is presented as digital twins \cite{el2018digital} in the virtual world. The parameters of physical devices can be collected by ubiquitous sensing technologies to maintain the same states as their corresponding digital twins. This is an interdisciplinary area that should cover lots of related subjects, such as material science, signal processing, internet of things (IoT), pattern recognition, etc \cite{essa2000ubiquitous, paulovich2018future}. On the contrary, after the operation and processing in metaverse, the parameters in virtual environments can be sent back to the physical devices and change their states in the real world. However, the development of digital twins about metaverse is still in its early stage, and there are the following research questions that deserve further study: \textit{(1) What kind of devices in the physical world should be conducted as digital twins in the metaverse? (2) How to use the digital twins in the metaverse to effectively benefit the real world?}
%(3) What kind of ubiquitous sensing technologies can maintain the real-time and persistent digital twins?} 

\textbf{Content creation interface.} The metaverse is a evolving virtual world with unlimited scalability and interoperability. The operators need to construct the basic elements, while innovative user-generated content (UGC) fulfill the universe through users. Therefore, high efficiency content creation is another significant component for interactions between users and the metaverse. For buildings, objects, and environments that exist in the physical world, we can apply 3D reconstruction approaches to build digital twins in the metaverse \cite{ma2018review}. To generate 3D models, users can utilize 3D modeling software, such as 3ds Max\footnote{https://www.autodesk.com/products/3ds-max/overview}, Blender\footnote{https://www.blender.org/}, or Maya\footnote{https://www.autodesk.com/products/maya/overview}, but these modeling systems are highly dependent on professional knowledge and experience, which is difficult for amateurs to replicate. So we summarize the following open research questions: \textit{(1) How to accurately reconstruct existing objects and their physical attributes in the virtual world? (2) How can the existing interaction modalities facilitate content creation to enhance the user experience?}

% \textbf{Content creation interface.} The metaverse is a virtual world with unlimited expansion, in which operators need to construct the basic elements, and users could arbitrarily create UGCs. Therefore, efficient content creation is another necessary component for the interaction between users and the metaverse. For the buildings, objects, and environments that exist in the physical world, we can apply 3D reconstruction approaches to build the digital twins in the metaverse \cite{ma2018review}. To generate 3D models, users can utilize 3D modeling software, such as 3ds MAX\footnote{https://www.autodesk.com/products/3ds-max/overview}, Blender\footnote{https://www.blender.org/}, or Maya\footnote{https://www.autodesk.com/products/maya/overview}, but these modeling systems highly depend on professional knowledge and experience, which are hard for amateurs. So we summarize the following open research questions: \textit{(1) How to accurately reconstruct the existing objects and their physical attributes in virtual world? (2) How can the existing interaction modalities facilitate the content creation to enhance the user experience?}
%(3) Are there any supplementary approaches that could assist the content creation, such as AI-based algorithms?}

\subsection{Ecosystem} \label{sec_ecosystem}

The ecosystem can provide a breathing and parallel living world that continuously serves all of the world’s inhabitants. Specifically, people may have social experiences that are completely different from the real world in activities, like befriending AI-driven non-player character (NPC). In this paper, we mainly discuss three parts consisting of the ecosystem layer: UGC, economics, and AI.

\textbf{User-generated content.} The UGC is any form of content that has been created by users rather than the developers/operators of online platforms \cite{krumm2008user, min2019blockchain}. Different from traditional game communities like Steam Workshop\footnote{https://steamcommunity.com/workshop/}, players’ goals are not to follow the developer’s rules, but rather to explore freely and create content in their own style. Hence, UGC in the metaverse tends to be heterogeneous and requires ownership. Blockchain-based Non-Fungible Token (NFT) provides a new approach to UGC in the metaverse, which could certify a digital asset to be unique and not interchangeable. Specifically, users can store their UGC as an NFT on the blockchain, and trade UGC through smart contracts to achieve liquidity. The free creation of UGCs can motivate the innovation of users, but it also brings some open research questions: \textit{(1) How to design more UGC-based applications to reflect and promote the value of UGCs? (2) How to design a reasonable mechanism to guarantee the uniqueness and reduce the malicious duplication of UGCs?}

\begin{figure*}[!t]
	\centering
	\includegraphics[width=2.1\columnwidth]{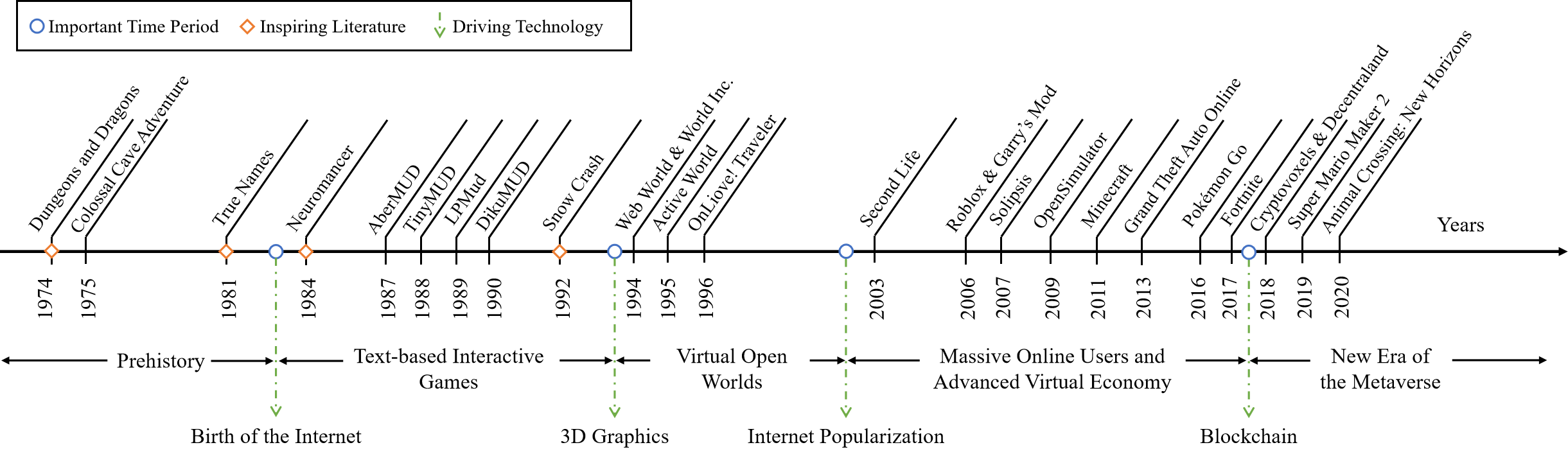}
	\caption{Brief Timeline of the Metaverse Development}
	\label{fig_timeline}
\end{figure*}

\textbf{Economics.} Economics is an important factor considered in the ecosystem, which could offer rich content and a vibrant community in the metaverse. Decentralized Finance, based on smart contracts and Fungible Token (FT), offers a way to innovate economic models in the metaverse. Existing successful solutions, such as Uniswap \cite{angeris2019analysis}, a Decentralized Exchange (DEX) on Ethereum, automatically provide users with liquidity for their tokens. At present, the main economic activities in the metaverse mainly include the auction of virtual assets, such as land, scarce items and precious real estate, development and leasing of land, rewards for finishing game tasks, and profits from investing in cryptocurrency. So the metaverse invokes a new form of funding that draws inspiration from both the real world and virtual world. However, the implementation of smart contract applications has the potential for external attacks, and the metaverse lacks decentralized financial applications. Hence, we summarize the research questions as follows: \textit{(1) How to design immutable yet sustainable smart contacts to maintain a balanced economic system for the metaverse? (2) How to design efficient DeFi models to improve the liquidity of NFTs in the metaverse?}

\textbf{Artificial intelligence.} AI facilitates our physical lives in many areas, including intelligent transportation \cite{pan2013trace}, smart healthcare \cite{varshney2007pervasive}, environmental monitoring \cite{ong2016dynamically}, and public safety \cite{liu2016deep}. Meanwhile, the key concept in the metaverse lies in its sophisticated data analytics for understanding, monitoring, regulating, and planning. Apart from the above-mentioned applications, AI-driven NPCs will be an indispensable part of our metaverse. Specifically, NPCs are computer-operated characters who act as enemies, partners, and support characters to provide challenges, offer assistance, and support the storyline. State-of-the-art AI mostly uses deep learning (DL) and reinforcement learning (RL), and achieves breakthrough progress in computer vision (CV) and natural language process (NLP). To give users a better user experience with NPCs, we formulate the following research questions: \textit{(1) How to facilitate users' operation using AI technology to enhance user experience in the metaverse? (2) What kind of AI technology can be used to effectively improve the comprehension and communication ability of NPCs?}

% \textbf{Artificial intelligence.} The AI facilitates our physical life in many areas like intelligent transportation \cite{pan2013trace}, smart healthcare \cite{varshney2007pervasive}, environment monitoring \cite{ong2016dynamically}, and public safety \cite{liu2016deep}. Meanwhile, the key concept in the metaverse lies on the sophisticated data analytic for understanding, monitoring, regulating, and planning. Apart from these applications in physical life, AI-driven NPCs will be an indispensable part of our Metaverse. Specifically, NPCs are computer-operated characters who act as enemies, partners, and support characters to provide challenges, offer assistance, and support the storyline. State-of-the-art AI mostly uses deep learning (DL) and reinforcement learning (RL), and achieves breakthrough progress in computer vision (CV) and natural language process (NLP). To give the users a better user experience with NPCs, we formulate the following research questions: \textit{(1) What kind of AI technology can be used to effectively improve the comprehension and communication ability of NPCs? (2) How to provide users a better interaction experience with AI-driven NPCs?}

\section{Journey Towards the Metaverse} \label{sec_related}

In fact, we are still on the way to exploring the metaverse, although this concept has been proposed for more than 30 years, as mentioned in Section \ref{sec_introduction}. Referring to fictional literature, such as \textit{True Names} \cite{vinge2001true}, \textit{Neuromancer} \cite{gibson2019neuromancer} and \textit{Snow Crash} \cite{stephenson2014snow}, they could help us picture a promising future about the metaverse. However, although there is no perfect example that meets all of the requirements of an ideal metaverse, various existing works possess several features that are worthy of summary. In this section, we will briefly introduce some remarkable metaverse forerunners, and illustrate a brief timeline, as shown in Figure \ref{fig_timeline}.

\begin{table*}[!t]
    \caption{Features of representative metaverse examples}
    \renewcommand{\arraystretch}{1.4}
    \begin{tabular}{cc|c|ccc|cc p{15pt}}
    \hline
         \multicolumn{2}{c|}{\multirow{2}*{Metaverse Examples}} & \multicolumn{1}{c|}{Infrastructure} & \multicolumn{3}{c|}{Interaction} & \multicolumn{3}{c}{Ecosystem} \\
    \cline{3-9}
         \multicolumn{2}{c|}{} & Blockchain & VR/AR & Digital twins & Creator & UGC & Economics & AI \\
    \hline
        Text-based Games & MUDs \& MUSHs & \textcolor{red}{\XSolidBrush} & \textcolor{red}{\XSolidBrush} & \textcolor{red}{\XSolidBrush} & \textcolor{green}{\Checkmark} & \textcolor{green}{\Checkmark} & \textcolor{red}{\XSolidBrush} & \textcolor{red}{\XSolidBrush} \\
        \hline
        \multirow{2}*{\makecell{Virtual Open \\ Worlds}} & Active World & \textcolor{red}{\XSolidBrush} & \textcolor{red}{\XSolidBrush} & \textcolor{red}{\XSolidBrush} & \textcolor{green}{\Checkmark} & \textcolor{green}{\Checkmark} & \textcolor{green}{\Checkmark} & \textcolor{green}{\Checkmark} \\
        
        & Solipsis & \textcolor{red}{\XSolidBrush} & \textcolor{red}{\XSolidBrush} & \textcolor{red}{\XSolidBrush} & \textcolor{green}{\Checkmark} & \textcolor{green}{\Checkmark} & \textcolor{red}{\XSolidBrush} & \textcolor{red}{\XSolidBrush} \\
        \hline
        
        \multirow{8}*{\makecell{MMO Video \\ Games}} & Second Life & \textcolor{red}{\XSolidBrush} & \textcolor{red}{\XSolidBrush} & \textcolor{red}{\XSolidBrush} & \textcolor{green}{\Checkmark} & \textcolor{green}{\Checkmark} & \textcolor{green}{\Checkmark} & \textcolor{green}{\Checkmark} \\
        
        & Roblox & \textcolor{red}{\XSolidBrush} & \textcolor{green}{\Checkmark} & \textcolor{red}{\XSolidBrush} & \textcolor{green}{\Checkmark} & \textcolor{green}{\Checkmark} & \textcolor{green}{\Checkmark} & \textcolor{green}{\Checkmark} \\
        
        & Minecraft & \textcolor{red}{\XSolidBrush} & \textcolor{green}{\Checkmark} & \textcolor{red}{\XSolidBrush} & \textcolor{green}{\Checkmark} & \textcolor{green}{\Checkmark} & \textcolor{green}{\Checkmark} & \textcolor{green}{\Checkmark} \\
        
        & Grand Theft Auto Online & \textcolor{red}{\XSolidBrush} & \textcolor{red}{\XSolidBrush} & \textcolor{green}{\Checkmark} & \textcolor{green}{\Checkmark} & \textcolor{green}{\Checkmark} & \textcolor{green}{\Checkmark} & \textcolor{green}{\Checkmark} \\
        
        & Pokémon Go & \textcolor{red}{\XSolidBrush} & \textcolor{green}{\Checkmark} & \textcolor{green}{\Checkmark} & \textcolor{red}{\XSolidBrush} & \textcolor{red}{\XSolidBrush} & \textcolor{green}{\Checkmark} & \textcolor{green}{\Checkmark} \\
        
        & Super Mario Maker 2 & \textcolor{red}{\XSolidBrush} & \textcolor{red}{\XSolidBrush} & \textcolor{red}{\XSolidBrush} & \textcolor{green}{\Checkmark} & \textcolor{green}{\Checkmark} & \textcolor{green}{\Checkmark} & \textcolor{green}{\Checkmark} \\
        
        & Fortnite & \textcolor{red}{\XSolidBrush} & \textcolor{red}{\XSolidBrush} & \textcolor{red}{\XSolidBrush} & \textcolor{green}{\Checkmark} & \textcolor{green}{\Checkmark} & \textcolor{green}{\Checkmark} & \textcolor{green}{\Checkmark} \\
        
        & Animal Crossing: New Horizons & \textcolor{red}{\XSolidBrush} & \textcolor{red}{\XSolidBrush} & \textcolor{red}{\XSolidBrush} & \textcolor{green}{\Checkmark} & \textcolor{green}{\Checkmark} & \textcolor{green}{\Checkmark} & \textcolor{green}{\Checkmark} \\
        \hline
        
        \multirow{2}*{\makecell{Decentralized \\ Virtual World}} & Cryptovoxels & \textcolor{green}{\Checkmark} & \textcolor{red}{\XSolidBrush} & \textcolor{red}{\XSolidBrush} & \textcolor{green}{\Checkmark} & \textcolor{green}{\Checkmark} & \textcolor{green}{\Checkmark} & \textcolor{green}{\Checkmark} \\
        
        & Decentraland & \textcolor{green}{\Checkmark} & \textcolor{red}{\XSolidBrush} & \textcolor{red}{\XSolidBrush} & \textcolor{green}{\Checkmark} & \textcolor{green}{\Checkmark} & \textcolor{green}{\Checkmark} & \textcolor{green}{\Checkmark} \\
    \hline
    \label{tab_feature}
    \end{tabular}
\end{table*}

\subsection{Pioneering Work}

\subsubsection{Text-based Interactive Games}

Text-based interactive game is the primary category of pioneering metaverse \cite{dionisio20133d}, which uses rules like the famous board game \textit{Dungeons \& Dragons} for reference. The typical examples are \textit{MUDs} (originally multi-user dungeon, with later variants, multi-user dimension and multi-user domain \cite{bartle2003designing, hahn1996internet}) and \textit{MUSHs} (Multi-User Shared Hallucination) \cite{shah1995playing}. Text-based interactive games build an online platform where players can communicate in real-time through texts and play collaboratively. After that, \textit{MUDs} and \textit{MUSHs} gradually evolved into different versions, such as \textit{AberMUD}, \textit{TinyMUD}, \textit{LPMud} and \textit{DikuMUD} with more features. Among them, \textit{TinyMUD} allows the user to create a game world for other players to explore, which marks the appearance of the UGC. The above-mentioned works represent the beginning of building the metaverse. We have selected Colossal Cave Adventure as the representative prototype. \textbf{Colossal Cave Adventure, Between 1975 and 1977}: \textit{Colossal Cave Adventure} is a role-playing and rogue-like game where players can move and interact with items through text commands, which describes the environment in text and gives results for certain actions.

% Text-based interactive games are the primary category of pioneering metaverse \cite{dionisio20133d}, which use the rules like the famous board game \textit{Dungeons \& Dragons} for reference. The typical examples are \textit{MUDs} (originally multi-user dungeon, with later variants multi-user dimension and multi-user domain \cite{bartle2003designing, hahn1996internet}) and \textit{MUSHs} (Multi-User Shared Hallucination) \cite{shah1995playing}. Text-based interactive games build an online platform where players can communicate in real-time through texts and play collaboratively. After that, \textit{MUDs} and \textit{MUSHs} gradually evolved into different versions, such as \textit{AberMUD}, \textit{TinyMUD}, \textit{LPMud} and \textit{DikuMUD} with more features. Among them, \textit{TinyMUD} allows the user to create their game world for other players to explore which marks the appearance of the UGC. The above-mentioned works represent the beginning of building the metaverse. We select Colossal Cave Adventure as the representative prototype. \textbf{Colossal Cave Adventure, Between 1975 and 1977}: \textit{Colossal Cave Adventure} is a role-playing and roguelike game where players can move and interact with the items through text commands, which describe the environment in text and give results for certain actions.

\subsubsection{Virtual Open Worlds}

In the 1990s, with rapid advances in computational power and computer graphics, people were no longer satisfied with text-based interactions. Virtual worlds equipped with 3D graphics and more open-ended socialization appeared at that time. Those virtual worlds built an online 3D virtual environment where people can log in as their corresponding avatar, who can create and construct UGCs in these worlds. Communication methods are also enriched to both text and voice for people to better share their experiences. Here are some typical examples: \textbf{Web World, 1994; Worlds Inc., 1994; Active Worlds, 1995}: This is a series of virtual worlds with the development of graphic technologies from 2.5D to 3D. Thereinto, \textit{Active Worlds} allows users to travel among 3D virtual worlds and environments built by others.

% In the 1990s, with the fast advance of computational power and computer graphics, people no longer satisfied with text-based interaction. Virtual worlds equipped with 3D graphics and more open-ended socialization appeared at that time. Those virtual worlds built an online 3D virtual environment where people can log in as a their corresponding avatars who can create and construct UGCs in these worlds. Communication methods are also enriched to both text and voice for people to better share their experiences. Here are some typical examples: \textbf{Web World, 1994; Worlds Inc., 1994; Active Worlds, 1995}: This is a series of virtual worlds with the development of graphic technologies from 2.5D to 3D. Thereinto, \textit{Active Worlds} allows users to travel among the 3D virtual world and environments built by others. 

\subsection{Modern Prototypes of the Metaverse}

In this section, we will introduce some existing and promising works that have representative features of the metaverse. Compared to the pioneers, these works are more technologically sophisticated and involve a larger group of users. According to their characteristics, we divide them into two categories as follows.
% the Massive Multi-player Online Video Games and the New Online Virtual World.

\subsubsection{Massive Multiplayer Online Video Games}

Massive multi-player online (MMO) video games are probably the most popular version of metaverse today. These games enable a massive amount of people to interact with each other in highly sophisticated 3D environments with convenient communication, UGC creation, economy, VR/AR, and so on. We list some representative examples of MMO video games as follows. Note that others like \textit{Grand Theft Auto Online}, \textit{Pokemon Go}, \textit{Animal Crossing} and \textit{Super Mario Maker 2} are also considered but not listed, due to limited space. \textbf{Second Life, 2003}: \textit{Second Life} is an online virtual world with a large amount of UGCs, with its own virtual currency, the Linden Dollar, which can be exchanged with real-world currency. \textbf{Roblox, 2006}: \textit{Roblox} is a platform where players can play games created by others. In \textit{Roblox}, players can use Roblox Studio to create games and virtual items, such as clothes, body parts, and gears, which can be bought and sold. \textit{Roblox} also occasionally hosts virtual and real-life events. \textbf{Minecraft, 2011}: \textit{Minecraft} is a sandbox video game where players can interact with a fully modifiable 3D environment made by blocks and entities. In \textit{Minecraft}, players can build astonishing architectures and innovative game levels with one-cubic meter-sized blocks, so \textit{Minecraft} has a massive number of UGCs. Moreover, \textit{Minecraft} supports VR devices, which highly enriches the user experience. \textbf{Fortnite, 2017}: \textit{Fortnite} is an MMO shooter game developed by Epic Games. \textit{Fortnite} allows players to construct buildings, create custom islands, and discover featured islands created by the community, but some in-game items, such as skins, gestures, and dances, are provided by the operator, which players could purchase. More importantly, Epic Games has already hosted many large in-game events, such as grand vocal concerts in \textit{Fortnite}.

\subsubsection{Decentralized Virtual World}

Different from the MMO video games, the decentralized virtual worlds are supported by blockchain technology, which has a build-in economy with an impact on the real economy. We highlight some examples as follows: \textbf{Cryptovoxels\footnote{https://www.cryptovoxels.com/}, 2018}: \textit{Cryptovoxels} is a virtual world built on Ethereum, which consists of a city called Origin City, owned by the operator, and parcels, owned by individuals. Users with an Ethereum wallet can trade the parcels and UGCs with others, and the parcel owner can freely modify its blocks and features. It is worth mentioning that some artists are displaying and trading their artwork in \textit{Cryptovoxels} as NFT. \textbf{Decentraland, 2018}: \textit{Decentraland} is another virtual world powered by Ethereum. Users can use Ethereum to trade empty land parcels of \textit{Decentraland}, and the owners can call software development kits (SDKs) provided by Decentraland, to build social games and applications. More importantly, \textit{Decentraland} has a sustainable creator economic system, in which land, estates, avatars, wearables, and even names can be traded in its marketplace. Encouraged by incentive mechanisms, a large number of UGCs, such as scenes, artworks, challenges, and buildings, are created by users in \textit{Decentraland}, constructing a virtuous circle.

% Different from the MMO video games, the decentralized virtual worlds are supported by blockchain technology which has a build-in economy that has an impact on the real economy. We highlight some examples as follows: \textbf{Cryptovoxels\footnote{https://www.cryptovoxels.com/}, 2018}: \textit{Cryptovoxels} is a virtual world built on the Ethereum, which consists of a city called Origin City owned by the operator and parcels owned by individuals. Users with an Ethereum wallet can trade the parcels and UGCs with others, and the owner of the parcel can freely modify its blocks and features. It is worth mentioning that some artists are displaying and trading their artwork in \textit{Cryptovoxels} as NFT. \textbf{Decentraland, 2018}: \textit{Decentraland} is another virtual world powered by Ethereum. Users can use the Ethereum to trade empty land parcels of \textit{Decentraland}, and the owners can call software development kits (SDKs) provided by \textit{Decentraland} to build social games and applications. More importantly, \textit{Decentraland} has a sustainable creator economic system, in which land, estates, avatars, wearables, and even names can be traded in its marketplace. Encouraged by positive mechanisms, a large number of UGCs, such as scenes, artworks, challenges, and architecture, are created by users in Decentraland, building a virtuous circulation. 

\subsection{Vision}\label{sec_vision}

From the architecture of the metaverse defined in Section \ref{sec_architecture}, we summarize seven main features of a metaverse, including blockchain, VR/AR, digital twins, UGC creator (denoted by creator), UGC, economics and AI, as shown in Table \ref{tab_feature}. Note that we do not list computation, communication, and storage, since they are the fundamentals of a digital world. Then we characterize the features of representative metaverse examples according to their type. Drawn from the table, we find that, the UGC and its creator are the only feature for pioneering metaverse examples. With technology development, the ecosystem becomes richer, where economics and AI begin to appear in Virtual Open Worlds. Coming to the era of MMO Video Games, the metaverse examples almost regard the UGC, economics, and AI as essential components. On the other hand, the MMO Video Games also present innovation in interactive modalities. \textit{Roblox}, \textit{Minecraft} and \textit{Pokémon Go} introduce VR/AR to provide users with a more immersive user experience, while \textit{Grand Theft Auto Online} and \textit{Pokémon Go} utilize digital twins to provide interactive experience refer to the physical world. After the breakthrough of blockchain-related technologies, novel metaverse examples introduce the blockchain to maintain decentralization and autonomy. More importantly, the ecosystem innovations resulting from the introduction of blockchain will be a promising research area.

\section{Campus Metaverse Prototype} \label{sec_metaverse}

\begin{figure}[!h]
    \centering
    \includegraphics[width=\columnwidth]{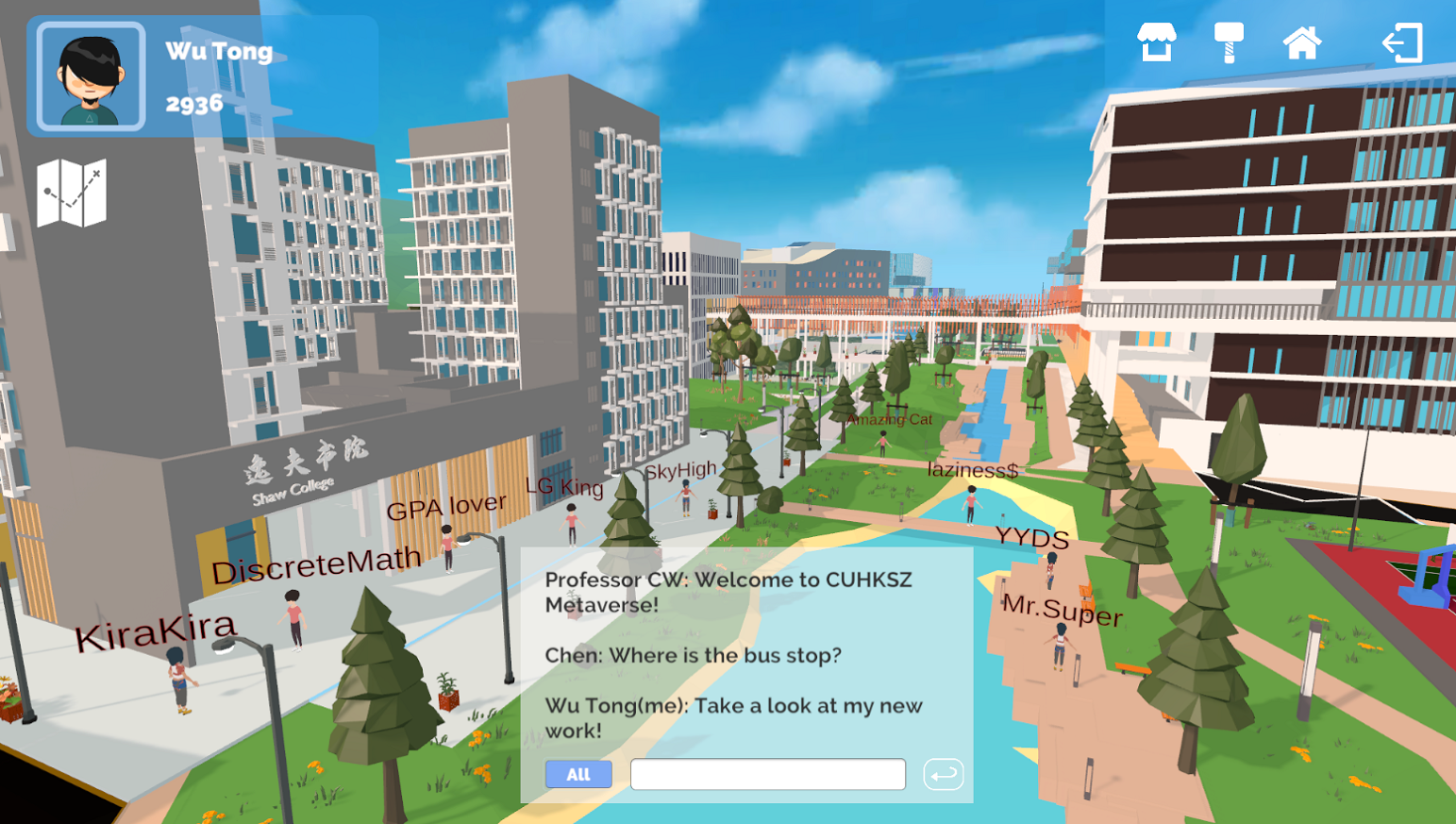}
  \caption{A corner at CUHKSZ through Metaverse Observer}
  \label{fig_observer}
\end{figure}

To achieve our goals and vision toward metaverse for social good, this section presents The Chinese University of Hong Kong, Shenzhen (CUHKSZ) Metaverse, our early-stage prototype, as a blockch-ain-driven exemplary system for demonstration and future social experiments. The proposed system aims to provide our on-campus students with an interactive metaverse, a mixed environment where students’ actions in the real world could correspondingly affect the virtual world, and vice versa. As shown in Figure \ref{fig_observer}, we illustrate a corner of CUHKSZ metaverse through the implemented Metaverse Observer. In the following subsections, we will introduce the key concepts of our system in detail.

% To make our goals and challenges towards metaverse for social good, this section presents The Chinese University of Hong Kong, Shenzhen (CUHKSZ) Metaverse, our early-stage prototype, as a blockchain-driven exemplary system for demonstration and future social experiments. The proposed system aims to provide our on-campus students an interactive metaverse, a mixed environment where the students' actions in the real world could correspondingly affect the virtual world, and vice versa. As shown in Figure \ref{fig_observer}, we illustrates a corner of CUHKSZ metaverse through the implemented Metaverse Observer. In the following subsections, we will introduce the key concepts of our system in detail.

\subsection{Infrastructure}

The proposed CUHKSZ Metaverse is developed on Unity\footnote{https://unity.com/}, a cross-platform development engine, thus the application can be deployed on smartphones, PCs, and browser-based cloud streaming, etc. With the support of the development platform, thousands of users from both cable and mobile network are all allowed to communicate and interact with each other in the CUHKSZ Metaverse. Further, the 3D models are constructed using Blender. More importantly, the CUHKSZ Metaverse applies blockchain with smart contract to support the ecosystem, including tokens, DAO, and trading system, etc. In the current version, we choose FISCO-BCOS\footnote{http://fisco-bcos.org/}, which is an open-source high-performance financial-grade consortium blockchain platform, to deploy CUHKSZ Metaverse. FISCO-BCOS can provide rich features including group architecture,  pluggable consensus mechanisms, privacy protection algorithms, etc. Also, Solidity\footnote{https://github.com/ethereum/solidity} is utilized as the smart contract programming language to support the blockchain-based ecosystem. The introduction of consortium blockchain in CUHKSZ Metaverse benefits the system operation and test in the early stages, and avoids the high cost of transactions in the public blockchain. In the next phase, we intend to connect the current consortium blockchain with the public blockchain, e.g., deploy and synchronize the data and smart contract to the public blockchain (such as Ethereum).

% The proposed CUHKSZ Metaverse is developed on Unity\footnote{https://unity.com/}, a cross-platform development engine, thus the application can be deployed in both smartphones, PCs, and browser-based cloud streaming. With the support of the development platform, the CUHKSZ Metaverse is built to afford the communication of thousands of users, where the connections from cable network, 4G and 5G mobile devices are all allowed. Besides, the 3D models are constructed using Blender. More importantly, the CUHKSZ Metaverse applies blockchain with smart contract to support the ecosystem, including the tokens, DAO, and trading system, etc. In current version, we choose FISCO-BCOS\footnote{http://fisco-bcos.org/}, which is an open-source high-performance financial-grade consortium blockchain platform, to deploy CUHKSZ Metaverse. FISCO-BCOS also provides an interface which can directly demonstrate the information of the consortium blockchain. Besides, the Solidity\footnote{https://github.com/ethereum/solidity} is utilized as the programming language of the smart contract to support the blockchain-based ecosystem. The introduction of consortium blockchain in CUHKSZ Metaverse is benefit for the system operation and test in early stage, which avoids the high cost of transactions in public blockchain. In the next phase, we intend to connect current consortium blockchain with public blockchain, e.g., deploy and synchronize the the data and smart contract to public blockchain (such as Ethereum).

\subsection{Interaction}

\subsubsection{Metaverse Viewer}

The Metaverse Viewer is built for users to interact with CUHKSZ Metaverse, which provides both a first-person and a third-person perspective, as shown in Figure \ref{fig_billboard}. To implement the Metaverse Viewer, we envision a cross-platform future that various devices would connect to CUHKSZ Metaverse, including smartphones, PCs, browser-based cloud streaming, etc. In the current prototype, we adopt the smartphone as the inaugural platform for the following reasons: (1) it is convenient to promote the metaverse concept to our target users, as most are heavy mobile phone users; (2) smartphones provide continuous access to the metaverse, especially in a mobile environment; (3) smartphones make it easy to acquire physical data from our users through sensors, cameras, and GPS modules built into the smartphone, which retains considerable potential for innovation with interactive approaches. However, the conventional smartphone is not yet to the ideal metaverse device. In the next phase, we envision providing an extension of a VR/AR interface to enhance the immersive user experience utilizing specific sensors, such as LiDAR.

% The Metaverse Viewer is built for users to interact with CUHKSZ Metaverse, which provides both first-person and third-person perspective, as shown in Figure \ref{fig_billboard}. To implement the Metaverse Viewer, we envision a cross-platform future that various devices would connect to CUHKSZ Metaverse, including smartphones, PCs, browser-based cloud streaming, etc. In the current prototype, we adopt smartphone as the inaugural platform for the following reasons: (1) it is convenient to promote the metaverse concept to our target users, as most of them are heavy mobile phone users; (2) smartphones provide all-time ubiquitous access to the metaverse, especially in a mobile environment; (3) smartphones make it easy to acquire physical data from our users through sensors, cameras, GPS modules equipped in the smartphone, which keeps much potential for innovation of interactive approaches. However, the conventional smartphone is yet to be the ideal device for the metaverse. In the next phase, we envision to provide the extension of VR/AR interface to enhance the immersive user experience utilizing specific sensors, such as LiDAR.

\subsubsection{Ubiquitous Sensing-based Service} \label{sec_lacation}

\begin{figure}[!b]
    \centering
    \begin{subfigure}[b]{0.47\linewidth}
    \centering
    \includegraphics[width=\textwidth]{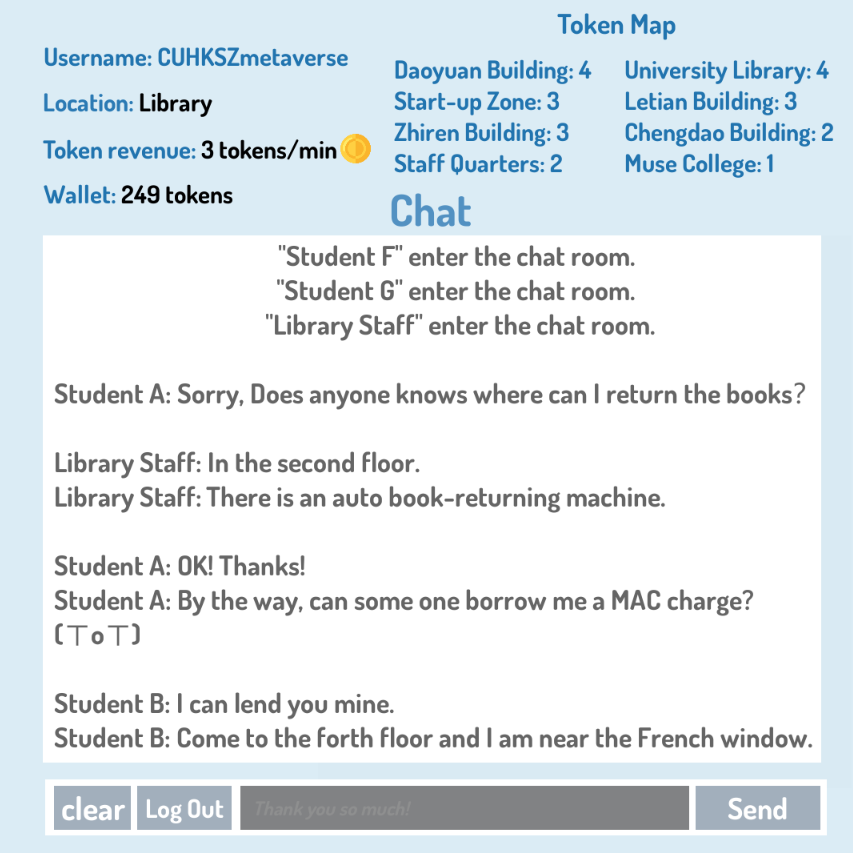}
      \caption{Location-based Services}
      \label{fig_chatting}
    \end{subfigure}
    \quad
  \begin{subfigure}[b]{0.47\linewidth}
    \centering
    \includegraphics[width=\textwidth]{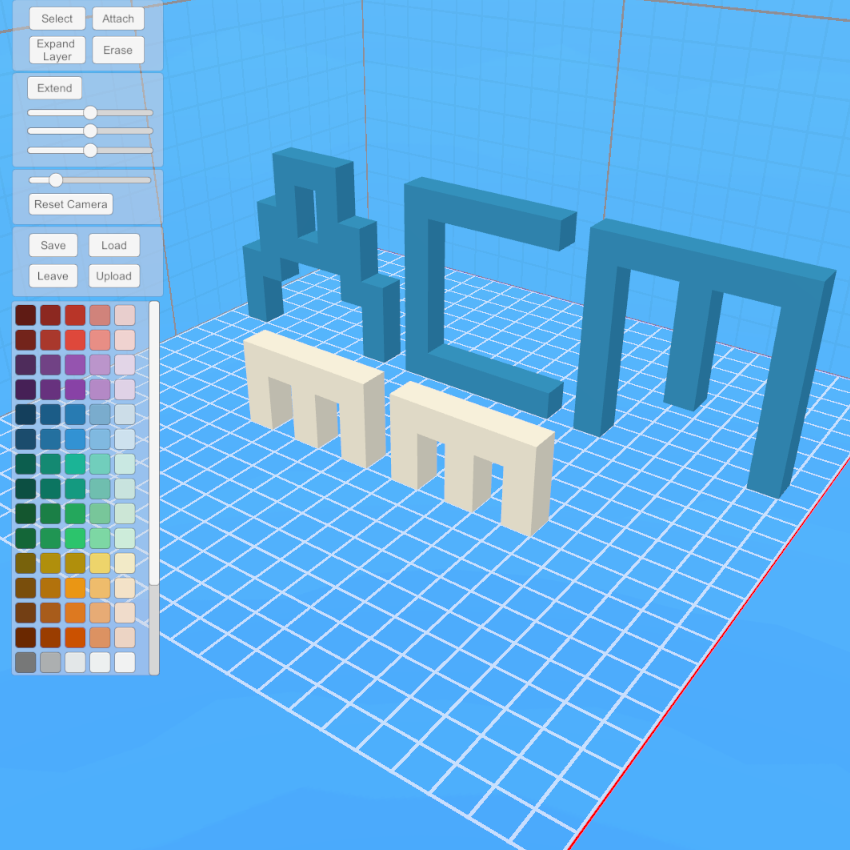}
      \caption{Editor for Content Creation}
      \label{fig_editor}
  \end{subfigure}
    \quad
  \caption{Interaction Layer in CUHKSZ Metaverse}
  \label{fig_ecosystem}
\end{figure}

Most of the existing metaverse enables interactions with traditional keyboards and mouses, which lack accurate control of the avatar and immersive user experience. The current version of our CUHKSZ Metaverse Viewer utilizes localization information as a source of sensing input. For instance, we feature a location-based incentive mechanism to maximize the social welfare of on-campus students: the students may start the Metaverse Viewer in Power-Saving mode and voluntarily report their GPS location for higher token production rates. As depicted in Figure \ref{fig_chatting}, the student is now physically studying in the University Library, thus, he/she will automatically join the University Library chat room to chat with nearby students through the corresponding metaverse channel, while earning tokens in high-speed mode. Apparently, this approach may encourage students to leave their dormitories and study at the library during the day. In the next phase of CUHKSZ Metaverse, we intend to leverage more sensing modalities, e.g. eye-tracking, to improve the user experience.

% Most of the existing metaverse enables interactions with traditional keyboards and mouses, which lacks the accurate control of avatar and immersive user experience. The current version of our CUHKSZ Metaverse Viewer utilizes the localization information as a source of sensing input. For instance, we feature a location-based incentive mechanism to maximize the social welfare of on-campus students: the students may start the Metaverse Viewer in Power-Saving mode and voluntarily report their GPS locations for higher token production rates. As depicted in Figure \ref{fig_chatting}, the student is now physically studying in University Library, thus, he/she will automatically join the chatting room of University Library to chat with nearby students through the corresponding metaverse channel while earning tokens in a high-speed mode. Apparently, this approach may encourage students to leave their dormitories and study in the library during the daytime. In the next phase of CUHKSZ Metaverse, we intend to leverage more sensing modalities, e.g. eye-tracking and LiDAR, to improve the users' quality of experience. 

\subsubsection{Content Creation} \label{sec_creator}

As envisioned in Section \ref{sec_vision}, UGC is a fundamental element in a metaverse. In CUHKSZ Metaverse, we also adopt UGC as the core user mechanism. However, it is still a challenge for an average user to create customized items in a 3D environment, as 3D object quality relies heavily on professional modeling knowledge and experience, as described in Section \ref{sec_interaction}. To address this issue, we designed and implemented an easy-to-use UGC editing tool, as shown in Figure \ref{fig_editor}. With this editor, an untrained user can learn to make their imaginary 3D items with voxels in minutes, as in creating architectures in Minecraft. Then the UGC editing tool will apply AI-based technologies to assist the UGC creation process, e.g., using generative algorithms to generate 3D items based on the voxels made by users. Moreover, the 3D items could automatically be transformed to a low-poly style model, which better fits the artistic style of the CUHKSZ Metaverse. Afterward, the created voxel model will be built as an NFT for more usage (e.g., trading and collection). In the next phase of CUHKSZ Metaverse, we intend to expand the UGC scope, e.g., allow users to create pets with AI-driven actions and emotions.

% As envisioned in Section \ref{sec_vision}, UGC is a fundamental element in a metaverse. In CUHKSZ metaverse, we also adopt UGC as the core mechanic for the users. However, it is yet a challenge for a normal user to create customized items in a 3D environment, as the quality of 3D objects heavily relies on professional modeling knowledge and experience as described in Section \ref{sec_interaction}. To address this issue, we design and implement an easy-to-use UGC editing tool, as shown in Figure \ref{fig_editor}. With this editor, an untrained user can learn to make their imaginary 3D items with voxels in minutes, as like creating architectures in Minecraft. Then the UGC editing tool will apply AI-based technologies to assist the creation process of users, e.g., using generative algorithms to generate 3D items based on the voxels made by users. Moreover, the 3D items could be automatically transformed to a low-poly style model, which better fits the artistic style of the CUHKSZ Metaverse. Afterward, the created voxel model will be built as an NFT for more usage (e.g., trading and collection). In the next phase of CUHKSZ Metaverse, we intend to expand the scope of UGC, e.g., allow the users to create pets with AI-driven actions.

\subsection{Ecosystem}

\subsubsection{Token-driven Ecosystem}

For a modern metaverse, a token-driven ecosystem is a key element. CUHKSZ metaverse employs blockchain-based tokens to feature a fair and transparent ecosystem, which are considered to be monetary representations for the community. In our current implementation, all residents can continuously claim tokens via smart contracts, which specify the number of tokens the users can collect in a certain period of time. According to the predefined rules, the token’s production rate may be varied for each user, subject to residents’ actions and performance in both the virtual and physical worlds, e.g. location-based incentive mechanism, which is discussed in \ref{sec_lacation}. The tokens can be utilized in various activities, such as trading in an official store, trading UGCs with other players, voting, etc.

% For a modern metaverse, the token-driven ecosystem is a key element. CUHKSZ metaverse employs blockchain-based tokens to feature a fair and transparent ecosystem, which are considered as the monetary representations for the community. In our current implementation, all residents can continuously claim tokens via smart contracts, which specify the number of tokens the user can collect in a certain period of time. According to the predefined rules, the token's production rate may be various for distinct users subject to the residents' actions and performance in both the virtual and physical worlds, e.g. location-based incentive mechanism, which will be discussed in \ref{sec_lacation}. The tokens can be utilized in various activities, such as trading in official store, trading UGCs with other players, voting, etc.

\begin{figure}[!b]
    \centering
    \begin{subfigure}[b]{0.47\linewidth}
    \centering
      \includegraphics[width=\textwidth]{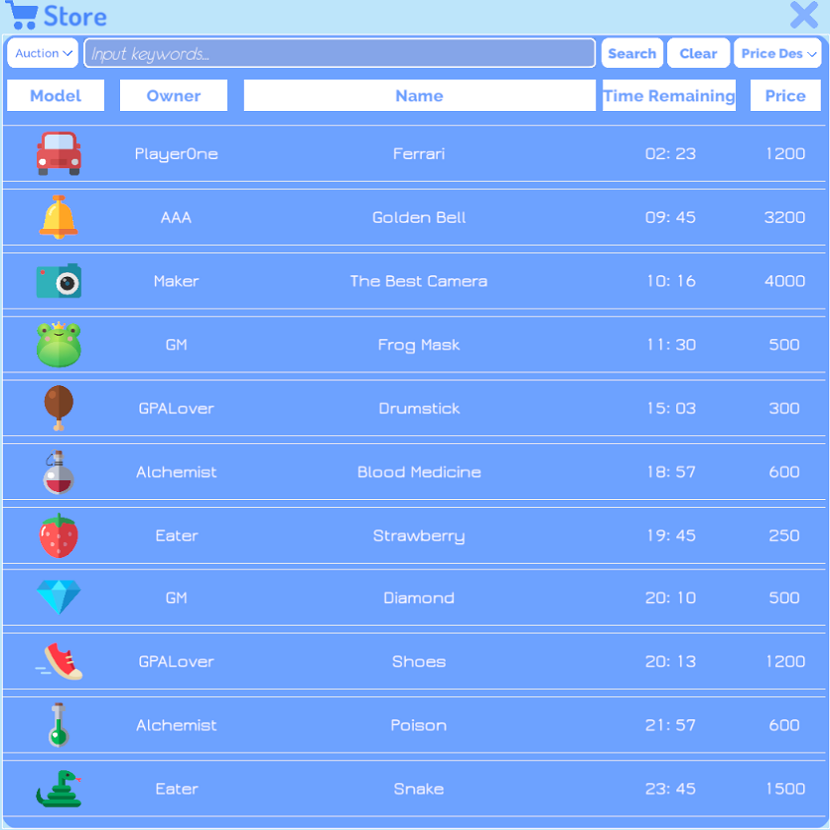}
      \caption{Trade and Purchase Services}
      \label{fig_store}
    \end{subfigure}
    \quad
    \begin{subfigure}[b]{0.47\linewidth}
    \centering
      \includegraphics[width=\textwidth]{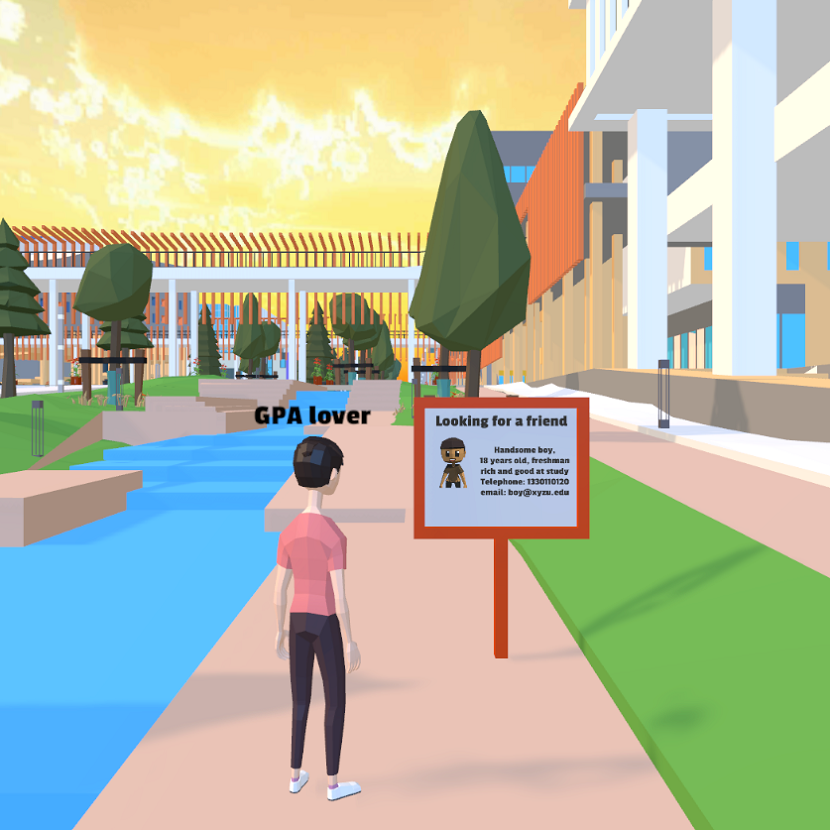}
      \caption{Billboard on Campus}
      \label{fig_billboard}
    \end{subfigure}
    \quad
  \caption{Ecosystem Layer in CUHKSZ Metaverse}
  \label{fig_ecosystem}
\end{figure}

\subsubsection{Autonomous Governance}

As discussed in Section \ref{sec_equality}, a well-designed democratic system may improve equality for the social good. CUHKSZ Metaverse enforces autonomous governance by introducing a Delegated Proof of Stake (DPoS) voting protocol for any motions or proposals to revise predefined rules, as described in Section \ref{sec_ecosystem}. According to the characteristics of an university, CUHKSZ Metaverse encourages students to establish a student union and elect a virtual committee, which could efficiently make decisions for matters representing users. For convenient discussion, we have also implemented an online forum to build the user community. This series of work assures autonomous governance and maintains the normal operation of society in the CUHKSZ Metaverse. Moreover, the iteration of CUHKSZ Metaverse is also based on the proposals of autonomous governance.

% As discussed in Section \ref{sec_equality}, a well-designed democratic system may improve equality for the social good. CUHKSZ Metaverse enforces autonomous governance by introducing a Delegated Proof of Stake (DPoS) voting protocol for any motions or proposals to revise predefined rules as described in Section \ref{sec_ecosystem}. Due to the property of the university, CUHKSZ Metaverse encourages students to establish the student union and elect the committee in the virtual world, which could efficiently make decisions for some matters representing other users. For convenient discussion, we also implement an online forum to build the community for users. This series of work assure autonomous governance and maintain the normal operation of the society in CUHKSZ Metaverse. Moreover, the iteration of CUHKSZ Metaverse is also based on the proposals of the autonomous governance.

\subsubsection{UGC Display and Trading}

In CUHKSZ Metaverse, we provide a UGC creator that allows users to create almost anything, which is discussed in \ref{sec_creator}. In the current stage, metaverse residents can trade UGCs with others, wear them as ornaments, or display them in their personal rooms. In addition, the billboard system provides a channel for users to give full play to their personalities in public. As illustrated in Figure \ref{fig_billboard}, a user may rent a certain area in the virtual campus and set their customized billboard, which may display their information or links to their personal room.

% In CUHKSZ Metaverse, we provide a UGC creator that allows users to create almost everything, which is discussed in \ref{sec_creator}. In the current stage, the metaverse residents can trade the UGCs with others, wear them as ornaments, or display them in their personal rooms. In addition, the billboard system provides a channel for the users to give full play to their personalities in public. As illustrated in Figure \ref{fig_billboard}, a user may rent a certain area in the virtual campus and set their customized billboard, which may display their information or links to their personal rooms. 

\subsubsection{AI-driven Metaverse Observer}

Different from the Metaverse Viewer, to observe and track key events of CUHKSZ Metaverse, we develop an AI-driven Metaverse Observer, as shown in Figure \ref{fig_observer}. This Metaverse Observer has a wider and higher vision that could cover an area of CUHKSZ Metaverse. Moreover, by tracking and analyzing real-time operation data from the Metaverse, this AI-driven Metaverse Observer can automatically recommend ongoing intriguing events to users, in which events with a high flow of users or approved by officials are more likely to be recommended. Therefore, the Metaverse Observer can provide global information for users and audiences, to better capture timely events.

% Different from the Metaverse Viewer, to observe and track key events of CUHKSZ Metaverse, we develop an AI-driven Metaverse Observer as shown in Figure \ref{fig_observer}. This Metaverse Observer has a wider and higher vision that could cover an area of CUHKSZ Metaverse. Moreover, by tracking and analyzing real-time operation data from the Metaverse, this AI-driven Metaverse Observer can automatically recommend ongoing intriguing events to users, in which events with a high flow of users or proved by officials are more likely to be recommended. Therefore, the Metaverse Observer can provide global information for users and audience to better capture the happening events.

\section{Conclusions} \label{sec_conc}

In this paper, we propose a three-layer architecture for metaverse, including infrastructure, interaction, and ecosystem, and the key components in each layer are discussed in detail, where a set of open research questions are summarized for each component. Afterward, we journey through the development of metaverse, describing both pioneering work and the novel metaverse examples. For intuitive illustration, a timeline of notable metaverse examples is drawn based on their release time, and a table is concluded to show the features of the representative metaverse. Lastly, the implemented blockchain-driven university campus prototype, CUHKSZ Metaverse, is illustrated, which could effectively enrich the campus life of university students and university faculties.

% In this paper, we propose a three-layer architecture for metaverse, including ecosystem, interaction, and infrastructure, and discuss the key components in detail for each layer to summarize a set of open research questions for each component. In Section \ref{sec_related}, we journey the development of metaverse containing both pioneering work and the novel metaverse. For intuitive illustration, we draw the timeline based on the release time of notable metaverse examples and conclude a table to show the features of the representative metaverse. At last, we illustrate the CUHKSZ metaverse prototype, which is a blockchain-driven virtual society that could enrich the campus life of university students and faculties. 

According to the representative applications, the metaverse significantly reflects the vision of human-centered computing, which is highly beneficial for the society in terms of accessibility, diversity, equality, and humanity. In the future, we will continuously improve our prototype and continue conducting metaverse-related studies. Not limited to the proposed questions, there are many research topics in the metaverse that are imperative to be studied. The core motivation of this paper is to direct more attention to the metaverse, and together, make a better society.

% According to the representative applications shown in Section \ref{sec_social}, the metaverse significantly reflect the vision of human-centered computing, which is highly beneficial for society in terms of accessibility, diversity, equality, and humanity. In the future, we will continuously improve our prototype and keep conducting metaverse-related studies. Not limited to the proposed questions, there are many research topics that are imperative to be studied in the metaverse. The core motivation of this paper is to raise more attention about metaverse, and together to make a better society.

%%
%% The acknowledgments section is defined using the "acks" environment (and NOT an unnumbered section). This ensures the proper identification of the section in the article metadata, and the consistent spelling of the heading.
\begin{acks}
This paper is supported in part by Project 61902333 from National Natural Science Foundation of China, and Project AC01202101038 from Shenzhen Institute of Artificial Intelligence and Robotics for Society. Thanks for other developers of CUHKSZ Metaverse: Honghao Chen, Nanjun Yao, Qiuhong Chen, Tong Chen, Xiangyu Xu, Yifan Zhao, Yiyan Hu, Yuyang Liang, Zexin Lin, Zhen Ren.
\end{acks}

%% The next two lines define the bibliography style to be used, and the bibliography file.
\balance
\bibliographystyle{ACM-Reference-Format}
\bibliography{reference}

\end{document}